\documentclass[12pt]{article}

\usepackage{amssymb}
\usepackage{amsmath}
\usepackage{amsfonts}
\usepackage{graphicx}
\usepackage{mathrsfs}
\usepackage{comment}
\usepackage{cite}

\begin{document}

\renewcommand{\figurename}{Fig.}
\renewcommand{\tablename}{Table.}
\newcommand{\Slash}[1]{{\ooalign{\hfil#1\hfil\crcr\raise.167ex\hbox{/}}}}
\newcommand{\bra}[1]{ \langle {#1} | }
\newcommand{\ket}[1]{ | {#1} \rangle }
\newcommand{\bef}{\begin{figure}}  \newcommand{\eef}{\end{figure}}
\newcommand{\bec}{\begin{center}}  \newcommand{\eec}{\end{center}}
\newcommand{\laq}[1]{\label{eq:#1}}  
\newcommand{\dd}[1]{{d \o d{#1}}}
\newcommand{\Eq}[1]{Eq.(\ref{eq:#1})}
\newcommand{\Eqs}[1]{Eqs.(\ref{eq:#1})}
\newcommand{\eq}[1]{(\ref{eq:#1})}
\newcommand{\Sec}[1]{Sec.\ref{chap:#1}}
\newcommand{\ab}[1]{\left|{#1}\right|}
\newcommand{\vev}[1]{ \left\langle {#1} \right\rangle }
\newcommand{\bs}[1]{ {\boldsymbol {#1}} }
\newcommand{\lac}[1]{\label{chap:#1}}
\newcommand{\SU}[1]{{\rm SU{#1} } }
\newcommand{\SO}[1]{{\rm SO{#1}} }
\def\({\left(}
\def\){\right)}
\def\dt{{d \o dt}}
\def\diag{\mathop{\rm diag}\nolimits}
\def\Spin{\mathop{\rm Spin}}
\def\O{\mathcal{O}}
\def\U{\mathop{\rm U}}
\def\Sp{\mathop{\rm Sp}}
\def\SL{\mathop{\rm SL}}
\def\tr{\mathop{\rm tr}}
\newcommand{\OR}{~{\rm or}~}
\newcommand{\AND}{~{\rm and}~}
\newcommand{\EV}{ {\rm ~eV} }
\newcommand{\KEV}{ {\rm ~keV} }
\newcommand{\MEV}{ {\rm ~MeV} }
\newcommand{\GEV}{ {\rm ~GeV} }
\newcommand{\TEV}{ {\rm ~TeV} }
\def\o{\over}
\def\a{\alpha}
\def\b{\beta}
\def\c{\varepsilon}
\def\d{\delta}
\def\e{\epsilon}
\def\f{\phi}
\def\g{\gamma}
\def\h{\theta}
\def\k{\kappa}
\def\l{\lambda}
\def\m{\mu}
\def\n{\nu}
\def\p{\psi}
\def\q{\partial}
\def\r{\rho}
\def\s{\sigma}
\def\t{\tau}
\def\u{\upsilon}
\def\v{\varphi}
\def\w{\omega}
\def\x{\xi}
\def\y{\eta}
\def\z{\zeta}
\def\D{\Delta}
\def\G{\Gamma}
\def\H{\Theta}
\def\L{\Lambda}
\def\F{\Phi}
\def\P{\Psi}
\def\S{\Sigma}
\def\me{\mathrm e}
\def\ol{\overline}
\def\tl{\tilde}
\def\*{\dagger}

\begin{center}

\vspace{1.5cm}

{\Large\bf Charge Quantization and Neutrino Mass\\ from Planck-scale SUSY }

\vspace{1.5cm}

{ \bf 
    Wen Yin\footnote{wyin@ihep.ac.cn}}

\vspace{12pt}
\vspace{1.5cm}
{\em 
  Institute of High Energy Physics,\\
\em  Chinese Academy of Sciences, Beijing 100049,  China \vspace{5pt}}

\vspace{1.5cm}
\abstract{

We show a possibility for the charge quantization of the standard model (SM) particles. 
If a global symmetry makes the three copies of a generation and supersymmetry (SUSY) relates the Higgs boson to a lepton, all the charges of the SM particles can be quantized through gauge-anomaly cancellation. 
In the minimal model realizing the possibility, the gravitino mass around the Planck-scale is needed to generate the SM couplings through (quantum) supergravity. 
Much below the Planck-scale, the SM with non-vanishing neutrino masses is obtained as the effective theory.
As extensions of the SM with this quantization mechanism, millicharged particles can exist without introducing massless hidden photons.
}
\clearpage
\end{center}
\setcounter{footnote}{0}
\section{Introduction}
The success of the standard model (SM) and standard big-bang cosmology clearly shows that there are new physics revealing the mysteries of  the origin of the neutrino mass, the identity of the dark matter, baryon asymmetry of the Universe, and the electric charge (charge) quantization of the visible particles.
The charge quantization can be explained by assuming a grand unified theory at a high energy scale~\cite{Georgi:1974sy} where the SM gauge group of $\SU(3)_c \times \SU(2)_L \times \U(1)_Y$ is embedded into a large compact gauge group. Alternatively, it was shown recently that charge quantization can be explained in a matter-coupled nonlinear sigma model~\cite{Hellerman:2013vxa}.

In this Letter, we show another simple explanation of the charge quantization of the SM particles.
It was pointed out that with only one generation the fermions must have 
quantized hypercharge to cancel quantum gauge anomaly~\cite{Geng:1988pr,
Minahan:1989vd}. However, the hypercharge of a scalar, e.g. the Higgs field, is not restricted.\footnote{By assuming non-vanishing fermion masses, the charge quantization for the Higgs field and also neutrinos were discussed in Refs. \cite{Babu:1989tq,
Babu:1989ex,Geng:1990nh}, but we will not make this assumption.} 
Our key idea is to relate the hypercharge of the Higgs boson to a lepton by supersymmetry (SUSY) 
and makes three copies of the generations of the SM by a global flavor symmetry. 
As a result, 
 all the SM particles' hypercharges are quantized from the gauge-anomaly cancellation.
This leads to the charge quantization when $\SU(2)_L\times \U(1)_Y$ is broken down.

We provide a SUSY model and find that at low energy the SM couplings, which may be forbidden by the symmetries for the mechanism, can be generated without introducing new superfields, thanks to (quantum) supergravity. 
It is believed that quantum gravity breaks any continuous global symmetry, and there could be various Planck-scale suppressed operators violating the flavor symmetry.
If the gravitino mass is around the Planck scale, $M_{\rm pl}\simeq 2.4 \times 10^{18}\GEV$, supergravity effect reduces the higher-dimensional operators into renormalizable ones with large enough coefficients, even the SUSY breaking sector is sequestered from the SM sector. As a result, the SM Lagrangian parameters can be generated.
Interestingly, if gaugino masses are generated dominantly by the anomaly mediation\cite{Giudice:1998xp, Randall:1998uk},\footnote{For recent studies on anomaly mediation in the MSSM, see Refs. \cite{Yin:2016shg,
Yanagida:2016kag,
Yanagida:2018eho}. } one of the observed neutrino oscillation scales can be explained by the seesaw mechanism~\cite{Yanagida:1979as, GellMann:1980vs, Minkowski:1977sc}.
 The extensions of the model and the mechanism are discussed.

\section{Charge quantization from supersymmetry}
It was known that with only one generation of the SM fermions, $X=Q,u,d,L,e$,  
 the hypercharges are guaranteed to be quantized from theoretical requirements~\cite{Geng:1988pr,
Minahan:1989vd}.
The quantum gauge field theory must satisfy the cancellation for chiral gauge and mixed gravity-gauge anomalies,
\begin{align}
2Y_L^3+Y_e^3+6Y_Q^3+3Y_u^3+3Y_d^3&=0 ~~~(\U(1)_Y^3),\nonumber\\
Y_L+3Y_Q&=0~~~(\U(1)_Y \SU(2)_L^2), \nonumber \\
2Y_Q-Y_u-Y_d&=0~~~(\U(1)_Y \SU(3)_c^2),\nonumber\\
2Y_L+Y_e+6Y_Q+3Y_u+3Y_d&=0~~~(\U(1)_Y {\rm gravity}^2). \laq{ac}
\end{align}
Here $Y_X$ is the hypercharge of a chiral fermion $X$, and the bracket denotes the corresponding triangle anomaly. 
If all the fermions are charged under $\U(1)_Y$~(= with non-vanishing hypercharge), the hypercharges are quantized as in the SM.\footnote{There is another solution with vanishing-hypercharge fermions, 
but we do not consider here.}

However, the hypercharge, $Y_H$, of the Higgs boson, $H$, does not contribute to the gauge anomaly. 
There may be no fundamental requirement for the Higgs boson to give masses to the fermions since a Yang-Mills theory without a Yukawa interaction or a mass term is still well-defined. 
Thus, throughout this Letter, we do not restrict the charge assignment of fields by requiring the existence of the operators in the Lagrangian. 
In more detail, we assume that under $\U(1)_Y$ a particle can be either a singlet\footnote{No singlet chiral multiplet under $\U(1)_Y$ will be introduced in the minimal scenario. } (hypercharge$=0$), or arbitrarily charged\footnote{Strictly speaking, when a hypercharge is not quantized, the gauge group is not compact, and thus is not an ``$\U(1)$" symmetry, but we will denote $\U(1)_Y$ for illustrative purpose.} 
(hypercharge$=$ real number not chosen by hand) with satisfying the gauge-anomaly cancellation and symmetries.
From this viewpoint, the charge quantization of the Higgs boson is non-trivial, and important for the existence of the Yukawa interactions and the fermion masses.

We stress that the charge of the Higgs boson can be quantized if there is SUSY where the Higgs boson is a (anti-)slepton.
This is because the hypercharge of the Higgs field becomes the opposite to that of the left-handed lepton,
 \begin{equation}
 Y_H=-Y_{L}.
 \end{equation}

Now consider a realistic case with three-generations, $X_\a=Q_\a,u_\a,d_\a,L_\a,e_\a$, where $\a=1,2,3$ represent the generations in a general basis.
The anomaly cancellation still restricts the hypercharge to be quantized if each type of the fermions carries the same hypercharge in all generations,
\begin{equation}Y_{X_\a}=Y_{X}.\end{equation}
This can be achieved from a global flavor symmetry, e.g. a global $\SU(3)$ symmetry where $\a$ is the index of the fundamental representation.

To sum up, if the SM is an effective theory of SUSY where the Higgs boson is a slepton with a global flavor symmetry, the charge quantization of all the SM particles is explained. 
The symmetries for this mechanism can be broken down at a very high energy scale, and the property of the symmetry may not appear at experimental scales except for the 
charge quantization. This is the main claim of the Letter.

Note that in the discussion, we have assumed no additional fermions that contribute to the anomalies in \Eq{ac}. 
We will go beyond this assumption in the last section.

\section{A SUSY extension of the SM}

Now we move to a particular scenario to relate a SUSY model to the 
SM. The hypothesis that the Higgs boson is a slepton
was originally pointed out in Ref.~\cite{Fayet:1976et}. 
After the Higgs boson discovery, this has been studied in the context of the SUSY phenomenology and was found to lead to various interesting signatures and better agreements with particular phenomena~\cite{Riva:2012hz,Bertuzzo:2012su, Berger:2015qra,Biggio:2016sdu}. 
For low-scale SUSY, $R$-symmetry with introducing adjoint multiplets is needed to avoid too large neutrino masses, and a low cutoff scale is needed to get realistic SM Yukawa interaction terms unless there are additional Higgs multiplets.

The major difference here is that we consider a scenario with a very high SUSY breaking scale instead of introducing new multiplets for simplicity. 
In particular, we do not introduce any fields to break the flavor and supersymmetries, which are needed for the charge quantization, in the SM sector, but suppose that the spontaneously SUSY breaking is in a sequestered sector~\cite{Inoue:1991rk, Randall:1998uk}
and this makes the cosmological constant vanishingly small.
Nevertheless, the symmetries can be badly broken and the couplings of the SM are generated by (quantum) supergravity, which mediates the SUSY breaking.
Using this scenario, we demonstrate that the slepton=Higgs hypothesis can lead to the charge quantization of the SM particles.
\\

The minimal model with the mechanism has the chiral supermultiplets 
\begin{equation}\F= Q_\a,u_\a,d_\a,L_\a,e_\a\end{equation} (we denote the corresponding fermion by itself, $\F$, and scalar by $\tl{\F}$) and vector superfields \begin{equation}V_i\end{equation} (corresponding field strength is ${\cal W}_a^{(i)}$), but no Higgs multiplets of the ordinary minimal supersymmetric standard model (MSSM). Here $i=Y,2,3$ is the index representing the SM gauge group.
For instance, let us identify the left-handed slepton of $\a=3$ as the Higgs field,
\begin{equation}
H\equiv \tl{L}^\*_3.
\end{equation}
Also, we assume there is a global $\SU(3)$ flavor symmetry where $\a$ is the index of the fundamental representation.
Accordingly, from the previous section, the charges of all the SM particles are quantized.

The component Lagrangian is given by 
\begin{equation}
{\cal L}= \int{d^4 \h  {\cal K}} +\int{d^2\h W} +h.c.+...,
\end{equation}
where the first (second) term represents a K\"{a}hler (super) potential. 
 $...$ represents the kinetic terms for vector multiplets, sequestered SUSY breaking Lagrangian, and the constant term for the superpotential which gives the gravitino mass $m_{3/2}.$
 It is believed that any continuous global symmetry is broken by quantum gravity. This implies that in the K\"{a}hler and superpotentials there are various Planck-scale suppressed operators which preserve the gauge symmetry but break the flavor symmetry. As we will see, these are the operators that generate most of the SM Lagrangian parameters through SUSY breaking, and the relevant operators will be soon discussed.

\paragraph{SM Yukawa matrices}
Thanks to the quantization of hypercharge, the SM Yukawa interaction terms are allowed to be written down in the low-energy effective Lagrangian.
We show that the Planck-scale suppressed operators due to quantum gravity would lead to general forms of the SM Yukawa matrices through supergravity.

One can write down the renormalizable superpotential,
\begin{equation}
\laq{W}
W\supset -\({1\over 2}{Y^{\a\b\g}_e} L_\a L_\b e_\g+Y^{\a\b\g}_d L_\a Q_\b d_\g+{1\over 2}{Y^{\a\b\g}_u } d_\a d_\b u_\g\),
\end{equation}
where  $Y^{\a\b\g}_{e,d,u}$ are Yukawa couplings in the SUSY model.
With the $\SU(3)$ symmetry, $Y_e^{\a\b\g}=Y^0_e \e_{\a\b\g}, Y^{\a\b\g}_{d}=Y^0_d \e_{\a\b\g},Y_u^{\a\b\g}=Y^0_u \e_{\a\b\g}$.
In general, in the K\"{a}hler potential, there is
\begin{equation} {\mathcal K}\supset -{1\over M_{\rm pl}} \({1\over 2}{\tl{Y}^{\a\b\g}_e} L_\a L_\b e_\g+\tl{Y}^{\a\b\g}_d L_\a Q_\b d_\g+{1\over 
2}{\tl{Y}^{\a\b\g}_u } d_\a d_\b u_\g\),\end{equation} 
where the dimensionless coefficients, $\tl{Y}_{u,d,e}$, are naturally non-zero and do not respect the flavor symmetry due to quantum gravity.
By performing a K\"{a}hler transformation, these terms are transfered into the superpotential as $Y_e^{\a\b\g}={m_{3/2}\over M_{\rm pl}}\tl{Y}^{\a\b\g}_e+Y^0_e \e_{\a\b\g}, Y^{\a\b\g}_{d}={m_{3/2}\over M_{\rm pl}}\tl{Y}^{\a\b\g}_d+Y^0_d \e_{\a\b\g},Y_u^{\a\b\g}={m_{3/2}\over M_{\rm pl}}\tl{Y}^{\a\b\g}_u+Y^0_u \e_{\a\b\g} $. 
After the transformation,  $\tl{Y}_{u,d,e}=0$ and ${Y}^{\a\b\g}_{u,d,e}$ are in general forms with the global symmetry $\SU(3)$ badly broken for $m_{3/2}\sim M_{\rm pl}$. 
From now on, we will work on the basis where $\tl{Y}_{u,d,e}=0$ for simplicity of the discussion.

The SM Yukawa matrices of leptons and down-type quarks have contributions from \Eq{W} as 
\begin{equation}
\laq{Yd}
(Y^{SM}_{e})^{\a\b}\supset {1\over 2}\(Y_{e}^{3\a\b}-Y_{e}^{\a3\b}\),\AND~(Y^{SM}_{d})^{\a\b}\supset Y_{d}^{3\a\b},
\end{equation}
 respectively, where the component Lagrangian we would like to get is
\begin{equation}
{\cal L}\supset -(Y^{SM}_e)^{\a\b} H^\* L_\a e_\b-(Y^{SM}_d)^{\a\b} H^\* Q_\a d_\b-(Y^{SM}_u)^{\a\b} H Q_\a u_\b.
\end{equation}
$Y_u^{SM}$ is the SM Yukawa matrix for up-type quarks which will be discussed soon.
Due to the contraction of the $\SU(2)_L$ indices, the first contribution in \Eq{Yd} is antisymmetric.

The matrix, $Y^{SM}_u$, and the symmetric part of $\({Y^{SM}_e}\)^{\a\b}$ are generated due to supergravity.  
To take into account the supergravity effect, let us follow a conformal compensator formulation of supergravity Lagrangian~\cite{Siegel:1978mj, Kugo:1982cu, Kugo:1983mv}. 
A chiral compensator multiplet, $\f$, is introduced to properly couple to superfields so that the action becomes superconformal invariant. Then we couple the action to superconformal gravity.
The Poiancar\'{e} supergravity can be obtained through the spontaneous breakdown of the local conformal symmetry. 
This is achieved from the non-vanishing ``vacuum expectation value (VEV)" of the compensator field, $\vev{\f}$. 
From the unitary gauge fixing, \begin{equation}\vev{\f} \equiv1+F_\f \h^2=\f-\d F_\f \h^2 ,\end{equation} and one gets supergravity Lagrangian by integrating out $\d F_\f$. 
With SUSY breaking, the $F$-term of $\vev{\f}$ is
 \begin{equation}F_\f = m_{3/2}\end{equation} 
 in most of the supergravity models.

Let us review the couplings among the compensator and superfields. 
A superconformal theory has dilatation and $\U(1)_A$ symmetries, under which a supermultiplet can have a Weyl wight $w$ and a chiral weight, $n$, respectively.
Thus the supermultiplet is characterized by $(w, n)$.
The chiral multiplet, $\f \OR \F=Q_\a,u_\a,d_\a,L_\a,e_\a$, has weights $(1,1)$ and the superfield strength, ${\cal W}_a^{(i)}$, has $(3/2,3/2)$. 
The covariant derivative of a $(0,0)$ multiplet, e.g. ${\mathcal D}_a (\F \f^{-1})$, carries $(1/2,-3/2)$. 
Notice that the covariant derivative(s) of an arbitrary supermultiplet is not necessary covariant under the dilatation and $\U(1)_A$ transformations. 
Proper introductions of the compensator field and decomposition of the spinor indices are needed for the covariance.
We do not discuss this in detail but refer to Ref.~\cite{Kugo:1983mv} when this is needed.
The conjugate of $(w,n)$ multiplet carries $(w,-n)$.
The weights of the supermultiplets should be added up to $(3,3)$ in the superpotential and $(2,0)$ in the K\"{a}hler potential by multiplying/dividing compensators and their conjugates.

Now let us focus on the relevant terms for $Y^{SM}_u \AND Y^{SM}_e$ by explicitly writing down the compensator dependence, 
\begin{equation}
\laq{Yknh}
{\mathcal K}\supset
-{  \tl{c}^{\a\b\g} \over M_{\rm pl}^2} \({\f \over \f ^\* }\)^2 L_{\a}^{\*}{\mathcal D}^a\(\f^{-1}u_\b\) {\mathcal D}_a \({ \f^{-1} Q_\g }\)-{ \tl{d}^{\a\b\g} \over M^2_{\rm pl}} {\f \over \f^{\*}} {\cal D}^a(\f^{-1} L_\a)  {\cal D}_a (\f^{-1} L_\b)  e_\g,
\end{equation}
where $\tl{c}^{\a\b\g}\AND \tl{d}^{\a\b\g}$ are dimensionless coefficients.
By substituting ``VEV" of $\f$, one obtains
\begin{equation}
\laq{Yu}
\(Y^{SM}_u\)^{\a\b}\supset8\times {\tl{c}^{3\a\b}}{\ab{F_\f}^2  \over M_{\rm pl}^2},~ \(Y^{SM}_e\)^{\a\b}\supset \(\tl{d}^{3\a \b}+\tl{d}^{\a3 \b}\)  {\ab{F_\f}^2 \over M_{\rm pl}^2}.
\end{equation}
Thus we found contributions to $Y^{SM}_u$ and the symmetric part of $Y^{SM}_e$.
There also exist other Planck-scale suppressed terms contributing to the SM Yukawa 
interaction,
which are suppressed (by $\O(F_\f/M_{\rm pl})$ or loop factors) compared to the effects already discussed. 
We do not discuss the other Planck-scale suppressed terms further, because with the other contributions given we can still get the SM Yukawa couplings by correctly choosing the parameters ${Y}_{e,d,u}^{\a\b\g},\tl{c},\tl{d}\lesssim \O(1), m_{3/2}\lesssim M_{\rm pl}$.

So far we have shown that the SM Yukawa interaction terms can be generated through (quantum) supergravity, even without a field to break the $\SU(3)$ flavor and supersymmetries in the SM sector. To have a realistic size of SM Yukawa couplings, especially for the top quark\footnote{The top quark Yukawa coupling is $y_t=\O(0.1)$ at around the Planck scale.}, 
\begin{equation}F_\f\simeq m_{3/2} \sim \O(0.1-1) M_{\rm pl} \laq{Ym32}\end{equation} 
is suggested if $\tl{Y}_{e,d,u}$ (before the K\"{a}hlar transformation),$\tl{c},\tl{d}\lesssim \O(1)$.\footnote{We do not consider $m_{3/2}>M_{\rm pl}$ because the cutoff scale is supposed to be around $M_{\rm pl}.$
If the cut-off scale is lower than $M_{\rm pl}$ (some of $\tl{Y}_{e,d,u},\tl{c},\tl{d}$ can be $>\O(1)$), the favored $m_{3/2}$ is smaller.}

\paragraph{Higgs field self-coupling} 
The Higgs field self-coupling in the SM has the contributions from $D$-term potentials,
Planck-scale suppressed terms such as 
${\cal K}\supset {1 \over M_{\rm pl}^2 \ab{\f}^2}\ab{L_3}^4,
$
and radiative corrections from certain mass parameters for the sparticles. (See c.f. Ref.~\cite{Vega:2015fna} for the MSSM.\footnote{A trilinear term such as ${\cal L}\supset \tl{L}_\a \tl{Q}_\b \tl{u}_\g$ can be generated for example from \Eq{Yknh}. For the mass squares, see the following. } )

\paragraph{Scalar masses}

The mass squares, $m_{\tl{\F}}^2$, for the scalar field, $\tl{\F}$, have contributions from: Planck-scale suppressed operators, anomaly mediation, and threshold corrections.\footnote{We do not consider Fayet-Iliopoulos term for $\U(1)_Y$ from theoretical consistency of supergravity~\cite{Komargodski:2009pc}.}
The first one could be from,
${\cal K}\supset {\tl{e}_\F \over M_{\rm pl}^2}
\ab{ \f^{-2} {\cal D}_a \f^{2}{\cal D}^a \f^{-1} \F}^2 $~\cite{Kugo:1983mv}, which provides Planck scale-suppressed mass, 
 $m_{\tl{\F}}^2\supset 4 \tl{e}_\F m_{3/2}^4/M^2_{\rm pl}.$ The anomaly mediation contribution is a 2-loop effect given by 
$
m_{\tl{\F}}^2\supset {m_{3/2}^2\over 2}{ d \g_\F \over dt},
$
where $\g_\F$ is the anomalous dimension of $\F$ and
$t=\log{\m_R}$ is the renormalization scale. 
This can be understood by substituting the ``VEV" of $\f$ in the wave function renormalization ${\mathcal K}\supset Z_\F(\mu_R/\ab{\f}) \F^\*\F$, where $\g_\F ={1\over 2}{d\over dt} Z_\F$.
Notice that the notorious tachyonic slepton problem can be  
avoided except for the Higgs field, when the Yukawa couplings, $Y_e^{\a\b\g}$ ($\a,\b\neq3$) and $Y_d^{\a\b\g}$ ($\a\neq3$), that are not important for the SM, are typically $\O(1)$. 
The threshold corrections are also loop-suppressed effects.

By summing up these contributions, the slepton ($\neq$ Higgs) and squark mass squares can be positive and are 
\begin{equation}
\laq{mass}
m_{\tl{\F}\neq H^\*}^2\sim \max{[ \O( m_{3/2}/16\pi^2)^2, \O(\tl{e}_\F({ m_{3/2}^2/ M_{\rm pl}})^2) ]}\sim \(\O(0.01-1)m_{3/2}\)^2. 
\end{equation}
The upper bound is obtained for $m_{3/2}\simeq M_{\rm pl}$ and $\tl{e}_\F= \O(1)$. For $m_{3/2}\lesssim \O(0.01) M_{\rm pl}$ or/and $\tl{e}_\F\ll1$, we get the lower bound.

Unfortunately, in our scenario the mass squared for the Higgs field should be canceled 
among the three contributions to realize the weak scale, $\sim 100 \GEV$, by finely tuning 
a parameter. 
By neglecting the Planck-scale suppressed term, the anomaly contribution is $ m_H^2 \supset - \O({1 \over 16 \pi^2} m_{3/2})^2$. 
This becomes comparable to the contribution from the Planck-scale suppressed operator with
$ m_{3/2} \sim \O({1 \over 16 \pi^2 \sqrt{\tl{e}_{L_3}} })M_{\rm pl}= \O(0.01-1)M_{\rm pl}$ for $\tl{e}_{L_3} \simeq \O(0.0001-1)$. (See \Eq{mass}.)
Thus one can make Higgs boson mass to be $\sim 100\GEV$ by finely tuning $\tl{e}_H$ with given $m_{3/2}=\O(0.01-1)M_{\rm pl}$ and other parameters.
The fine-tuning means we can not solve or alleviate the hierarchy problem between the Planck and weak scales without other assumptions/mechanisms. 
A detailed approach to this problem is out of our purpose, but let us mention the possibility of anthropic principle~\cite{Agrawal:1997gf}, and a candidate of solution~\cite{Arkani-Hamed:2016rle}. 
In the last section, we will discuss the extension of this mechanism in terms of low energy SUSY.

We have shown that the masses of the sleptons and squarks are naturally around $\O(M_{\rm pl})$. As we will see in the next section, the gaugino masses are slightly smaller than $\O(M_{\rm pl})$.
This implies that when the Higgs boson mass is finely tuned to be at a small value, only the particles in the SM are light, which interact with each other through the SM couplings and higher dimensional operators. Consequently, the SM is obtained as the effective theory at the scale $\ll M_{\rm pl}$. 

Notice that the higher dimensional operators in the component Lagrangian are generated by integrating out the sparticles whose masses are much larger than PeV scale. 
This implies that the flavor and CP-violating processes generated by the sparticle propagation are highly-suppressed and hence the SUSY flavor and CP problems are solved.\footnote{The gravitino problem is also absent. The gravitino whose mass is around $M_{\rm pl}$ even does not appear in the thermal history of inflationary cosmology as long as the inflation scale is much smaller than $M_{\rm pl}$.}
However, a proton decay rate and neutrino masses may obtain significant contributions, since they are very sensitive to the UV physics that breaks lepton/baryon number. These will be discussed in the following.

\paragraph{Proton decay}
Our scenario does not have any global symmetry, and a proton decays at a tiny rate.
By integrating out the squarks, $\tl{d}_\g$ from the second and third terms of \Eq{W}, one gets the four Fermi interaction,
\begin{equation}
{\cal L}\supset C^{\a\b\d\e} L_\a Q_\b (u_\d d_\e)^*,
\end{equation}
where $C^{\a\b\d\e}\equiv \sum_{\g } {Y_d^{\a\b\g} (Y_{u}^{\g\d\e})^* \over m_{\tl{d}_\g}^2}$. This gives the dominant contribution.
The decay rate of $p\rightarrow \pi^0+e^+$ is given by~\cite{Aoki:2017puj}
\begin{equation}
{\G_p}^{-1} \sim 3\times 10^{37}~ {\rm yrs} \({ 1/\sqrt{C^{ (1111)_{SM}}} \over 0.01M_{\rm pl} }\)^4,
\end{equation}
where the supscript ``$(1111)_{SM}$'' represents that we have focused on the interaction among the fermions in the 1st generation of the SM.
Since ${ 1/\sqrt{C^{ (1111)_{SM}}}}\gtrsim \O(m_{\tl{d}_\g})$ for $Y^{\a\b\g}_d, Y^{\a\b\g}_e<\O(1)$, with \Eq{mass} and $m_{3/2}\simeq M_{\rm pl}$ the decay rate is well below the current bound, $\G_p < \G_p^{\rm bound}\simeq \(1.6\times 10^{34}~{\rm yrs}\)^{-1}$~\cite{Miura:2016krn}.%
~If $1/\sqrt{C^{(1111)_{SM}}} \lesssim \O(0.01) M_{\rm pl}$, 
our scenario can be tested in future experiments~(e.g. in Hyper-Kamiokande~\cite{Abe:2011ts}).

\section{Neutrino scale and gaugino mass}

The hypothesis, ``Higgs=slepton", introduces lepton-number violation~\cite{Fayet:1976et}. 
A non-vanishing neutrino mass is a direct consequence, where 
the neutralinos can play the role of the right-handed neutrino in the seesaw mechanism~\cite{Yanagida:1979as,
GellMann:1980vs, Minkowski:1977sc}. 

In particular, if a neutralino mass, $M_{i=Y,2}$, is around $(10^{-3}-10^{-2}) M_{\rm pl}$, the measured neutrino 
oscillation scale, $\sqrt{|\D m^2_\n|}\sim 10^{-2}-10^{-3}\EV$~\cite{Patrignani:2016xqp}, could appear 
from the seesaw relation $\sim v^2/M_i$, where $v=\vev{H}\simeq 174\GEV$. 
This is achieved if the gaugino mass is loop-suppressed, and thus particular attention should be paid to the possibility that the masses are generated through anomaly mediation in our scenario.

In what follows, we will assume that the anomaly mediation effect is dominant for the gaugino masses, which is justified in certain cases. In our scenario, the only other possibility to generate the gaugino mass is the higher derivative terms, such as~\cite{Kugo:1983mv}
${\cal K}\supset -{\tl{e}_i \over 4 M_{\rm pl}^2 \f \f^\*} \tr[\({\cal D}_{b}{(\f^{-2} {\mathcal W}^{(i)}_{a})}+{\cal D}_{a}{(\f^{-2} {\mathcal W}^{(i)}_{b})}\)^2]$, where $\tl{e}_i$ is a dimensionless parameter.\footnote{The dimension five term ${\mathcal K} \supset {1 \over M_{\rm pl}} ({\cal W}^{(i)})^2 $ is the gauge kinetic term with performing a K\"{a}hler transformation.}
The gaugino masses are obtained from $\vev{\f}$ as $ \tl{e}_i {m_{3/2}^3\over M_{\rm pl}^2}$. 
This would be subdominant with $\sqrt{\tl{e}_i}m_{3/2}<  \O(0.1) M_{\rm pl}$, which implies that
$m_{3/2}$ is slightly smaller than the Planck scale and/or $\tl{e}_i$ is small. We note that $\tl{e}_i$ may be suppressed because the terms including $>2$ derivatives with two fields give rise to ghost modes and are absent in ghost-free effective theories~\cite{Fujimori:2017kyi}.

The gaugino mass terms from anomaly mediation are given by~\cite{Giudice:1998xp, Randall:1998uk}
\begin{equation}
\laq{anmd}
M_i\simeq {\b_i \over g_i} F_\f.
\end{equation}
 By following Refs.~\cite{Martin:1993zk,
Yamada:1994id,
Jack:1994kd}, the beta-functions, $\beta_i$, for the gauge couplings $g_i$ are calculated as,
\begin{align}
\b_{Y}&\simeq {5 g_Y^3 \over 8\pi^2},\\
\b_{2}&\simeq \frac{g_2^3}{64 \pi ^4}{ \left(\frac{9 g_2^2}{2}+6 g_3^2+\frac{g_Y^2}{2}-2\sum_{\a\b\g}{\(3{\ab{Y_{d}^{\a\b\g}}^2+\ab{Y_{e}^{\a\b\g}}^2}\)}\right)},\\
\b_{3}&\simeq-\frac{3 g_3^3}{16 \pi ^2}
\end{align}
at the leading order where we have neglected the contribution from the Planck-scale suppressed terms.
Notice that we do not have the ordinary Higgs multiplets, and the beta functions for $g_Y \AND g_2$ differ from the MSSM ones. 
In particular, $g_2$ runs at the 2-loop level. Hence, the anomaly-induced mass for wino is two-loop suppressed from $F_\f=m_{3/2}$ and can give a
 contribution to the neutrino mass even for $m_{3/2}\simeq M_{\rm pl},$ as we will see soon.

The coupling of Higgs-neutrino-wino arises from the gauge interaction in
\begin{equation}
\laq{kin}
{\mathcal K}\supset L_3^\* L_3 .
\end{equation}
The relevant component Lagrangian are obtained as 
\begin{align}
{\mathcal L}\supset -{ g_2 \over \sqrt{2} }  H \l_2  L_3- {M_2\over2} \l_2 \l_2  +h.c..
\end{align}
Here $\l_2$ is the wino triplet.
By integrating out $\l_2$, an effective neutrino mass term appears as
\begin{equation}
{\cal L}\supset {m_{\n33}\over 2v^2} HL_3HL_3,~~ m_{\n33}={1\over 4}{v^2  \over  M_2}  g_2^2.
\end{equation}
The neutrino mass reads from \Eq{anmd}:
\begin{align}
\laq{nv2}
m_{\n33}&\simeq 4 \times 10^{-2}\EV \({M_{\rm pl}\over m_{3/2}}\)~~~~~~~~~~~~~( {g_i{~\rm dominant}}),\\
m_{\n33}&\simeq -6 \times 10^{-2}\EV \({\sum{\(3{\ab{Y_{d}}^2+\ab{Y_{e}}^2}\)} \over 
66 }\) \({0.01M_{\rm pl}\over  m_{3/2}}\)\nonumber \\ 
&~~~~~~~~~~~~~~~~~~~~~~~~~~~~~~~~~~~~~~~~~({Y_e, Y_d}~{\rm dominant }),
\end{align}
where we have assumed the Yukawa couplings that are not important for the SM are negligible (dominant) in the first (second) row and $ g_Y=0.6\sqrt{3/5}, g_2=0.5,g_3=0.5$. 

Therefore, one finds that for \begin{equation}F_\f=m_{3/2}=\O(0.01-1)M_{\rm pl},\end{equation} one of the neutrino mass appears around the observed scales of the neutrino oscillation.
This range of $m_{3/2}$ overlaps with the one for badly broken $\SU(3)$ flavor symmetry and realistic SM Yukawa couplings in \Eq{Ym32}.

\paragraph{Another neutrino mass}
To explain the neutrino oscillation, there should be at least one additional mass term for a different flavor of neutrino, say $L_{2}$.
One leading candidate to generate the mass is through the seesaw mechanism with bino, $\l_Y$, as the right-handed neutrino. 
However, the gauge interaction in \Eq{kin} is flavor-blind, and contribute to the mass of the same neutrino $L_3$. Accordingly, a flavor-breaking coupling from quantum gravity
through the terms, such as ${\mathcal K}\supset {1 \over M^2_{\rm pl}}
{1\over \f^\*} {\cal D}^a \(\f^{-1}L_\a \) {\cal W}_a^{(i)} L^\*_3$, is needed. 
In this case, a similar term for wino coupling should be also taken into account, but this does not change \Eq{nv2} significantly unless the coefficient is extremely large. 
Since the lightest neutrino is massless,
the effective Majorana neutrino mass, $m_{\n \rm ee}= \O(0.001) \EV$ and $\O(0.01) \EV $, is predicted for normal and inverted mass hierarchies, respectively. This gives non-vanishing rate for the neutrino-less double beta decay, where our scenario may be tested.

Alternatively, there could be an additional massive singlet superfield, $N$, who has a coupling of type ${\mathcal K}\supset \({\f \over \f ^\* }\)^2 {\cal D}_\a (\f^{-1}L_2) {{\cal D}^\a (\f^{-1}N)} L^\*_3$. This reduces to the Yukawa couplings to the right-handed neutrino, ${\cal L}\supset 8\times ({m_{3/2}\over M_{\rm pl}})^2 H L_2 N$. Hence one obtains another neutrino mass through the conventional seesaw mechanism. 

In both cases, if the reheating temperature of the Universe is sufficiently high, the baryon asymmetry can be explained through thermal leptogenesis~\cite{Fukugita:1986hr} or through leptogenesis via neutrino oscillation during the thermalization of the Universe~\cite{Hamada:2018epb}.

\section{Discussion and Conclusions}

So far we found that the SM particles can have quantized charge due to an extremely high scale SUSY with flavor symmetry. 
It would be interesting to study the SM Yukawa structure from a particular SUSY and flavor symmetry breaking scenario~(c.f. Ref.~\cite{Liu:2005rs}), which leads to the charge quantization. 
Our mechanism can also apply to certain extensions of the SM.
For example, one can identify several sleptons as multi-Higgs doublets, or introduce some $\U(1)_Y$ singlets coupled to the SM. 

The extensions with additional hypercharged multiplets have restrictions not to spoil the mechanism. 
For instance, one can add hypercharged chiral multiplets, $E$ and $\ol{E}$, of fundamental and anti-fundamental representations under a new $\SU(N)$ gauge 
symmetry, respectively. To cancel the $\SU(N)^2\U(1)_Y$ gauge anomaly, the hypercharges, $Y_E \AND Y_{\ol{E}}$, satisfy  $Y_E=-Y_{\ol{E}}$ (c.f. \Eqs{ac}). Since $E$ and $\ol{E}$ are forced to have vector-like representation, the charge quantization in the SM sector  is not affected.  

In particular, $Y_E=-Y_{\ol{E}}$ can be arbitrary, and thus the corresponding electric charge is not 
quantized in general,\footnote{The $\SU(N)$ can be broken down by a hidden Higgs (without hypercharge) or becomes strongly-coupled at low energy.} which implies that  
 there could be millicharged particles (MCPs). A particle without quantized charge is stable, and it can be (a fraction of) dark matter.\footnote{Alternatively the dark matter could be a QCD axion with the decay constant around the Planck scale if the inflation scale is low~\cite{Graham:2018jyp, Guth:2018hsa}. The inflaton itself could also be the dark matter~\cite{Kofman:1994rk,
Kofman:1997yn,
Daido:2017wwb,
Daido:2017tbr,
Hooper:2018buz}.} Interestingly, the 21-cm anomaly reported by EDGES Collaboration~\cite{Bowman:2018yin} can be explained by a fraction of a MCP dark matter~\cite{Barkana:2018qrx, Berlin:2018sjs, Munoz:2018pzp}. 
The leading mechanism to get MCPs has been known to be the kinetic mixing between the SM photon and a massless hidden photon~\cite{Holdom:1985ag}.
However, in the presence of a dark photon, the explanation of 21-cm anomaly may run afoul to the dark radiation constraints~\cite{Vogel:2013raa,Ade:2015xua}.  We stress that a MCP can exist without hidden photons if the quantization of the charges in the SM is due to gauge-anomaly cancellation.
In particular, our mechanism can provide an elementary MCP, and this possibility is free from the constraints for dark photons.

A similar mechanism for charge quantization can be considered in a low energy $N=1$ SUSY, e.g. the MSSM, originating from a broken extended SUSY at a high energy scale. 
For example, the up Higgs $H_u$ with $Z=L_\a, H_d$ (down Higgs) in the MSSM can form a hypermultiplet $(H_u, Z)$ in a $N=2$ SUSY theory, which means the hypercharges of $H_u$ and $Z$ are opposite.
The $N=2$ SUSY theory can be partially broken down to $N=1$ at a high energy scale~\cite{Bagger:1994vj}.
Then the vanishing conditions of the gauge anomaly require the hypercharge to be quantized with a certain global flavor symmetry, which may also relate $H_d$ and $L_\a$.
Through the $N=2$ gauge interaction $W\supset {g_Y \over \sqrt{2}} \f_Y L_3 H_u$ and the mass term, $W\supset {1\over 2}M_Y \f_Y \f_Y$, of the vector partner $\f_Y$, a neutrino mass is generated.
An $N=2$ non-renormalization theorem can lead to typical spectra for the MSSM sparticles~\cite{Yin:2016pkz,Shimizu:2015ara}.

In this Letter, we have shown that if the Higgs boson is a slepton and there is a global flavor symmetry at an extremely high-scale, the charge quantization for all the SM fields can be guaranteed from the cancellation of gauge anomaly.  
If the gravitino mass is around the Planck scale, the (quantum) supergravity effect leads to the SM with the flavor and supersymmetries badly broken. 
A consequence of the quantization mechanism is the non-vanishing neutrino mass. 
In particular, one of the observed neutrino scales can be explained through the anomaly-induced gaugino masses.

\section*{Acknowledgments}
We thank Ryuichiro Kitano, Satoshi Shirai, and Kazuya Yonekura for fruitful discussion.  In particular, we thank Ryo Yokokura for useful communication on the compensator formulation with covariant derivatives.

\end{document}